%% file: main.tex
\documentclass[runningheads,orivec]{llncs}
\usepackage[T1]{fontenc}
\usepackage{graphicx,verbatim}
\usepackage{amsmath}
\usepackage{amsfonts}
\usepackage{multirow}
\usepackage{booktabs}
\usepackage[shortcuts]{extdash}
\usepackage{tikz}
\usepackage{cancel}
\usepackage[normalem]{ulem}
\usepackage{hyperref}
\usepackage{xurl}
\usepackage{orcidlink}

\usepackage[table]{xcolor}

\hypersetup{
    colorlinks,
    linkcolor={red!50!black},
    citecolor={blue!50!black},
    urlcolor={blue!50!black}
}

\usepackage{color}

\begin{document}
\title{Flow Matching with Optimized Subclass Priors for Medical Image Augmentation}
\titlerunning{Flow Matching with Subclass Priors}
\begin{comment}  %% Removed for anonymized MICCAI submission
\author{First Author\inst{1}\orcidID{0000-1111-2222-3333} \and
Second Author\inst{2,3}\orcidID{1111-2222-3333-4444} \and
Third Author\inst{3}\orcidID{2222--3333-4444-5555}}
%
\authorrunning{F. Author et al.}
\institute{Princeton University, Princeton NJ 08544, USA \and
Springer Heidelberg, Tiergartenstr. 17, 69121 Heidelberg, Germany
\email{lncs@springer.com}\\
\url{http://www.springer.com/gp/computer-science/lncs} \and
ABC Institute, Rupert-Karls-University Heidelberg, Heidelberg, Germany\\
\email{\{abc,lncs\}@uni-heidelberg.de}}

\end{comment}

\author{Felix Nützel\inst{1}\orcidlink{0009-0000-8277-172X}
%index{Nützel, Felix}
\and
Mischa Dombrowski\inst{1}\orcidlink{0000-0003-1061-8990}
%index{Dombrowski, Mischa}
\and
Bernhard Kainz\inst{1,2}\orcidlink{0000-0002-7813-5023}
%index{Kainz, Bernhard}
}
\authorrunning{F. Nützel et al.}
\institute{Department of Artificial Intelligence in Biomedical Engineering, Friedrich-Alexander-Universität Erlangen-Nürnberg, 91052 Erlangen, Germany
\email{felix.nuetzel@fau.de}
\and
Department of Computing, Imperial College London, London SW7 2AZ, UK \\
}
  
\maketitle              % typeset the header of the contribution
\begin{abstract}
Rare diseases dominate the diagnostic challenge in medical imaging yet
are severely underrepresented in clinical datasets, causing classifiers
to fail on exactly the conditions where reliable detection matters most.
Generative augmentation can supply the missing tail-class coverage, but
coarse disease labels aggregate diverse subtypes and acquisition
settings into multi-modal conditionals that bias generators toward
dominant submodes, while a shared Gaussian source forces rare
subpopulations through disproportionately long transport paths.
We propose an offline strategy that introduces informative priors at two
levels: first, we partition each coarse label into coherent submodes via
Gaussian mixture modeling in the generative model's latent space;
second, we learn subclass-conditioned source distributions that
re-center and re-scale the starting distribution per submode, shortening
trajectories and reducing within-subclass dispersion.
To prevent degenerate solutions we impose explicit geometric control,
moderately concentrating normalized displacement directions around
learnable prototypes while capping path-length outliers.
On long-tailed chest X-ray (MIMIC-LT, NIH-LT) and CT slice (CT-RATE) benchmarks
the proposed method consistently improves tail-class generation fidelity
and diversity (FID, IRS) and is a promising augmentation strategy that
reliably improves downstream balanced accuracy and macro-F1 over a
non-augmented baseline across modalities. Code is available at \url{https://github.com/Felix-012/OptPriorFM}

% Authors must provide keywords and are not allowed to remove this Keyword section.

\end{abstract}
\section{Introduction}

The classification of medical images under severe long-tail imbalance remains a central obstacle in clinical machine learning. Real-world datasets in chest radiography~\cite{holste2022long,wang2017chestx}, computed tomography~\cite{hamamci2024foundation}, and other modalities are dominated by ``no finding'' or common conditions, while clinically urgent but easily overlooked findings such as small pneumothoraces or pneumomediastinum on chest radiographs and
bronchiectasis on CT occur at very low prevalence. Standard discriminative training therefore overfits to head classes and underrepresents rare findings, producing classifiers that perform worst precisely where diagnostic accuracy is most consequential~\cite{holste2022long}. Generative data augmentation can, in principle, supply the missing tail-class coverage, and diffusion-based synthesis has shown encouraging results in radiology~\cite{rajaraman2024addressing}, ultrasound~\cite{che2025subtyp}, and across histopathology and dermatology~\cite{ktena2024genmodfair}. In practice, however, conditional generators trained with coarse disease labels face two compounding difficulties. First, a single diagnostic label often aggregates diverse subtypes, acquisition protocols, and comorbidities into multi-modal conditionals that bias the generator toward dominant submodes~\cite{bao2022conditional,na2024labelnoise}, effectively truncating rare but clinically valid variation. Second, drawing every generation trajectory from a shared standard Gaussian source forces some subpopulations through disproportionately long transport paths, inflating regression difficulty and degrading tail fidelity~\cite{lipman2022flow,pooladian2023multi}. Recent analyses of source distribution design for flow matching confirm that the geometry of the learned transport strongly affects generation quality, but also that naive directional objectives can encourage degenerate behavior such as norm inflation or overly concentrated paths~\cite{issachar2025designing,lee2026is,kim2026bettersourcebetterflow}.

We address both issues with an offline strategy that introduces informative priors at two levels. For each coarse label~$c$ we fit a Gaussian mixture model in the generative model's latent space to induce hard subclass assignments~$k$, partitioning heterogeneous conditionals into coherent submodes with lower conditional variance. We show that this subclass conditioning cannot increase the Bayes-optimal flow-matching risk; the improvement equals the expected between-subclass variability of the conditional mean displacement (Eq.~\ref{eq:improvement_exact}). We then learn subclass-dependent source distributions $\pi_0(\cdot\mid c,k)$ that re-center and re-scale the starting distribution per subclass, shortening typical trajectories and reducing within-subclass dispersion. We control the geometry of the induced displacements by encouraging moderate angular compactness of normalized directions around learnable subclass prototypes while capping path-length outliers to prevent norm-inflation shortcuts. The underlying process of creating subclasses and increasing their alignment has a direct analogy in long-tailed representation learning~\cite{hou2023subclass}: our method effectively learns a better representation of the interpolation paths induced by the latent representation of the target.

\noindent\textbf{Contributions.}
\textbf{(i)}~We introduce a subclass-conditional formulation for long-tailed medical image generation that partitions multi-modal conditionals via GMM-induced hard assignments.
\textbf{(ii)}~We propose subclass-dependent learned sources as informative priors that reshape conditional transport and reduce within-subclass dispersion.
\textbf{(iii)}~We develop a practical proxy objective that encourages moderate angular compactness while explicitly controlling path-length outliers, preventing norm-inflation shortcuts and avoiding overly restrictive directional collapse.
\textbf{(iv)}~We empirically demonstrate improved tail coverage and downstream long-tail classification on chest X-ray and CT benchmarks.

\noindent\textbf{Related Work.}
Generative augmentation for imbalanced medical imaging has advanced rapidly, with diffusion-based synthesis improving classification in pediatric chest X-rays~\cite{rajaraman2024addressing}, breast ultrasound subtyping~\cite{che2025subtyp}, and histopathology and dermatology with demonstrated fairness gains~\cite{ktena2024genmodfair}; lesion-focused diffusion enables controllable pathology synthesis~\cite{zhang2025lefusion} and causal disentanglement targets long-tail robustness~\cite{montenegro2025causal}.
Flow matching~\cite{lipman2022flow} has emerged as an efficient alternative, further improved by minibatch optimal-transport couplings~\cite{pooladian2023multi}, with applications to defect classification~\cite{oh2025flaw} and reweighted long-tailed generation via unbalanced optimal transport~\cite{song2025reweighted}.
To improve long-tail generation, guidance strategies attempt to recover modes at sampling time~\cite{sehwag2022ghfid,um2024minority,morshed2025diverse,um2025selfguided} but cannot recover modes never learned by the model; training-time remedies include class-balanced objectives~\cite{qin2023cb}, conditioning strategies~\cite{dombrowski2025irs}, expert adapter modules~\cite{nuetzel2025grasp}, reinforcement learning~\cite{miao2024reinforce}, knowledge transfer~\cite{zhang2024longtailed}, and label-noise robust training~\cite{na2024labelnoise}. Refining conditioning by clustering or pseudo-labels has proven effective~\cite{adaloglou2025rethinking}, with hierarchical generative clustering exploiting latent tree structure~\cite{Goncales2026tree} and theoretical analyses confirming that partitioning multi-modal distributions into simpler components improves learnability~\cite{bao2022conditional}. However, none of these split existing coarse medical labels into directional subclasses. Separately, data-dependent priors accelerate convergence in audio synthesis~\cite{lee2022priorgrad}, conditional prior distributions map conditions to Gaussian mixture sources~\cite{issachar2025designing}, and recent work uses online condition-dependent source optimization~\cite{kim2026bettersourcebetterflow} and analyzes risks of overly concentrated source designs~\cite{lee2026is}. Our approach complements these perspectives by introducing subclass conditioning with geometry-regularized source design, tailored to long-tailed, multi-modal medical conditionals.

\section{Method}

\input{figures/method}
%\input{figures/balanced_geo}
%\input{tables/subclass}
%\input{tables/optimization_diagnostics}
\input{tables/diagnostics_unified}

An overview is shown in Fig.~\ref{fig:method}. In long-tailed regimes, a coarse condition $c$ typically hides a second layer of imbalance: samples within $\pi_1(\cdot\mid c)$ exhibit attribute-wise long tails in which frequent attributes dominate and rare attributes are sparsely observed. This within-condition skew biases learning toward dominant modes and truncates rare but semantically valid directions of variation. We address this at three levels: (i)~we refine coarse labels by inducing hard subclass assignments that partition heterogeneous conditionals into coherent submodes, (ii)~we learn subclass-dependent source distributions that re-center and re-scale the starting distribution per submode, and (iii)~we impose explicit geometric control on the resulting displacements to prevent degenerate solutions (see Fig.~\ref{fig:bal_geo}).

\noindent\textbf{Irreducible error in flow matching.}
Given a source $\pi_0$ and a class-conditional target $\pi_1(\cdot\mid c)$, linear conditional flow matching interpolates $x_t = (1{-}t)\,x_0 + t\,x_1$ with $t\sim\mathcal{U}[0,1]$ and fits a velocity field $v_\theta(x_t,t\mid c)$ by minimizing $\mathbb{E}\bigl[\|v_\theta(x_t,t\mid c) - d\|^2\bigr]$ where $d = x_1 - x_0$ is the per-sample displacement. The Bayes-optimal minimizer is $v^*(x,t\mid c) = \mathbb{E}[d \mid x_t{=}x, t, c]$, and the irreducible (Bayes-optimal) risk equals the expected conditional variance of $d$:
\begin{equation}
    \min_{v(\cdot)}\;
    \mathbb{E}\!\left[\|v(x_t,t\mid c) - d\|^2\right]
    \;=\;
    \mathbb{E}\!\left[\mathrm{Var}(d \mid x_t,t,c)\right].
    \label{eq:bayes-risk}
\end{equation}
Any mechanism that reduces $\mathrm{Var}(d\mid x_t,t,c)$ therefore directly lowers the irreducible regression error.

\noindent\textbf{Subclass conditioning.}
\label{sec:latent_space}
We refine $c$ by inducing a discrete subclass label $k$ in the VAE latent space used by the flow-matching model. We use the frozen FLUX1.dev VAE and compute deterministic latents $x_1$ via the encoder mean. For each class $c$ we compute the training-split class center $\mu_c = \mathbb{E}[x_1\mid c]$ and fit a diagonal Gaussian mixture model to the residual latents $\tilde{x}_1 := x_1 - \mu_c$, selecting the number of components $K_c$ by EBIC with $\gamma{=}0.5$ (Table~\ref{tab:subclass}). Each sample receives a hard assignment $k = a_\phi(x_1,c) := \arg\max_{j} q_\phi(j\mid \tilde{x}_1,c)$, inducing a partition $S_{c,k} = \{x : a_\phi(x,c) = k\}$ with empirical subclass distributions $\pi_1(x_1\mid c,k) := \pi_1(x_1\mid c,\, x_1 \in S_{c,k})$ and weights $\pi_1(k\mid c) := \mathbb{P}[x_1 \in S_{c,k}\mid c]$, so the class-conditional distribution decomposes exactly as $\pi_1(x_1\mid c) = \sum_{k=1}^{K_c} \pi_1(k\mid c)\,\pi_1(x_1\mid c,k)$. We then condition the velocity field on $(c,k)$ and allow the source to depend on the refined label via $\pi_0(\cdot\mid c,k)$, while retaining the linear interpolation $x_t = (1{-}t)\,x_0 + t\,x_1$.

Performing subclass induction and all source optimization directly in the same latent space as the flow-matching model ensures that the induced subclasses, the learned sources, and the resulting probability paths share the same geometry, while avoiding any split leakage in subclass discovery.

\noindent\textbf{Subclassing cannot increase the Bayes risk.}
Conditioning on the refined label $k$ cannot increase the Bayes-optimal risk. Let $A = (x_t, t, c)$ and $B = k$. The law of total variance gives
%\begin{equation}
$
    \mathrm{Var}(d \mid A)
    \;=\;
    \mathbb{E}\!\left[\mathrm{Var}(d \mid A,B)\,\middle|\,A\right]
    \;+\;
    \mathrm{Var}\!\left(\mathbb{E}[d \mid A,B]\,\middle|\,A\right).
    \label{eq:totvar_rewrite}
%\end{equation}
$
Taking expectations over $A$ and applying~\eqref{eq:bayes-risk} yields the exact improvement decomposition
\begin{equation}
    \mathbb{E}\!\left[\mathrm{Var}(d \mid x_t,t,c)\right]
    -
    \mathbb{E}\!\left[\mathrm{Var}(d \mid x_t,t,c,k)\right]
    =
    \mathbb{E}\!\left[
      \mathrm{Var}\!\left(\mathbb{E}[d\mid x_t,t,c,k] \,\middle|\, x_t,t,c\right)
    \right]
    \ge 0.
    \label{eq:improvement_exact}
\end{equation}
The improvement is strict whenever $k$ changes the conditional mean displacement on a set of nonzero measure, \emph{i.e.}, whenever $\mathbb{P}\!\bigl(\mathrm{Var}(\mathbb{E}[d\mid x_t,t,c,k]\mid x_t,t,c) > 0\bigr) > 0$. 
%\noindent\textbf{Inference-time subclass handling.}
At sampling time conditioned on class $c$, we draw $k \sim \hat{p}(k\mid c)$ from the empirical mixture weights, sample $x_0 \sim \pi_0(\cdot\mid c,k)$, and integrate the learned velocity field conditioned on $(c,k)$. This realizes the class-conditional generator as a mixture $p_\theta(x\mid c) = \sum_{k=1}^{K_c} \hat{p}(k\mid c)\,p_\theta(x\mid c,k)$, avoiding any need to infer $k$ from a target sample at test time.

\noindent\textbf{Subclass-dependent source optimization.}
Subclassing reduces ambiguity between modes, but within each $(c,k)$ the displacement $d = x_1 - x_0$ can still be highly dispersed and heavy-tailed in norm when a single global source is used, increasing irreducible regression difficulty and yielding outlier trajectories. We therefore learn a subclass-dependent source $\pi_0(\cdot\mid c,k) = \mathcal{N}(\mu_{c,k},\,\mathrm{diag}(\sigma_{c,k}^2))$ with learnable $\mu_{c,k}\in\mathbb{R}^D$ and $\sigma_{c,k}\in\mathbb{R}^D_{+}$ (optimizing $\log\sigma_{c,k}$), so that for $z\sim\mathcal{N}(0,I)$ we sample $x_0 = \mu_{c,k} + \sigma_{c,k}\odot z$.

Directly minimizing $\mathbb{E}[\mathrm{Var}(d\mid x_t,t,c,k)]$ with respect to the source parameters is intractable because $x_t = (1{-}t)\,x_0 + t\,x_1$ depends on $x_0$ and hence on the source itself. We therefore optimize an offline proxy on $d = x_1 - x_0$ that targets two goals: improved directional structure of normalized displacements and explicit control of path-length outliers. The sources are initialized at their empirical class means, if this lowers angular subclass overlap and reduces the radial variance. This is motivated by recent analyses showing that directional objectives can simplify transport but admit degenerate solutions when radial behavior is unconstrained~\cite{kim2026bettersourcebetterflow}.

\noindent\textbf{Directional compactness with fixed outlier caps.}
Write $d = r\,u$ with $r = \|d\|$ and $u = d/\|d\|$. We introduce a learnable unit prototype $v_{c,k}\in\mathbb{S}^{D-1}$ per subclass and encourage moderate angular compactness via $\mathcal{L}_{\mathrm{out}} = \mathbb{E}[1 - \langle u, v_{c,k}\rangle]$. A key failure mode of purely angular losses is that they can be trivially improved by increasing $r$ (norm inflation), since larger displacements make $u$ less sensitive to additive variability in $x_0$. To prevent this, we impose an explicit penalty on path-length outliers using a fixed, subclass-specific cap $r^{\mathrm{cap}}_{c,k} = Q_{0.99}(\|x_1 - \mu_c\|\mid c,k)$, the 99th percentile of distances to the class center among samples assigned to $(c,k)$, computed once and held constant during optimization:
$\mathcal{L}_{\mathrm{path}} = \mathbb{E}\!\bigl[\mathrm{softplus}(\|d\|/r^{\mathrm{cap}}_{c,k} - 1)^2\bigr]$.
Since $r^{\mathrm{cap}}_{c,k}$ is fixed, this loss cannot be circumvented by inflating the cap, serving as a guardrail against norm-inflation shortcuts. To prevent trivial collapse or explosion of the diagonal scales, we regularize log-scales toward a unit reference via $\mathcal{L}_{\mathrm{det}} = \mathbb{E}[\|\log\sigma_{c,k}\|_2^2]$. The final offline objective is
$\mathcal{L} = \lambda_{\mathrm{out}}\,\mathcal{L}_{\mathrm{out}} + \lambda_{\mathrm{path}}\,\mathcal{L}_{\mathrm{path}} + \lambda_{\mathrm{det}}\,\mathcal{L}_{\mathrm{det}}$,
which empirically reduces within-subclass dispersion and suppresses heavy-tailed trajectories (Tab.~\ref{tab:geom_diag}).

\section{Experiments}

\noindent\textbf{Datasets.}
We validate our method on two medical long-tailed benchmarks spanning different imaging modalities. For chest radiography, we use MIMIC-LT~\cite{holste2022long} and NIH-LT~\cite{holste2022long}, the single-label long-tailed variants of MIMIC-CXR~\cite{johnson2024mimic} and NIH-CXR~\cite{wang2017chestx}. For computed tomography, we extract axial slices from CT-RATE~\cite{hamamci2024foundation} restricted to single-pathology slices following~\cite{dombrowski2025lcmem}, yielding a heavily imbalanced dataset of 14 classes and 22{,}305 samples. We additionally confirm consistent improvements on a synthetic 2D benchmark. %(Fig.~\ref{fig:synth}). 
Evaluation of generative quality is provided by Fr\'{e}chet Inception Distance (FID) and generation diversity with IRS~\cite{dombrowski2025irs}, both computed using domain-specific DenseNet feature extractors (a CT-RATE slice classifier~\cite{dombrowski2026pso} and a CXR classifier~\cite{dombrowski2025lcmem}). Downstream classification is assessed by balanced accuracy (bAcc) and macro-F1.

\noindent\textbf{Implementation details.}
Source optimization runs for 2{,}500 steps with $\lambda_{\mathrm{out}}{=}1$, $\lambda_{\mathrm{path}}{=}\lambda_{\mathrm{det}}{=}0.1$. Each FLUX transformer flow-matching model (i.e., baselines, method, and ablations) was trained for 48\,h on 8$\times$H100 GPUs with three seeds. Subclass induction and source optimization are fully offline and take less than one GPU-hour per dataset, negligible relative to flow-matching training. Downstream ResNet-50 classifiers were trained with three seeds per generative seed (nine seeds total per method), with early stopping after 15 epochs without validation improvement. Generative metrics compare 50{,}000 synthetic samples against training splits. All baselines use the same FLUX architecture, optimization settings, and a learned class-token conditioning. We compare against vanilla FM, CBFM~\cite{qin2023cb}, and CPD~\cite{issachar2025designing} as the most directly comparable offline source-modification baseline.

We use targeted class balancing based on training-split frequencies.  $n_c$ is the count of class $c$, $c^\star$ the dominant class, and $m = \mathrm{median}(\{n_c : c \neq c^\star\})$. Classes are partitioned into ULT ($n < m/4$), LT ($m/4 \le n < m/2$), MT ($m/2 \le n < 5m/2$), and Head ($n \ge 5m/2$). Only ULT/LT/MT classes are augmented with targets $t_{\text{ULT}} = \min(\lceil 0.5m \rceil, 10n)$, $t_{\text{LT}} = \min(\lceil 1.0m \rceil, 10n)$, $t_{\text{MT}} = \min(\lceil 2.5m \rceil, 5n)$; Head classes and \texttt{No~Finding} are unchanged. This yields 9{,}094 synthetic samples for MIMIC-LT, 5{,}454 samples for NIH-LT, and 2{,}784 for CT-RATE.

\input{tables/table4-7}

\input{figures/qualitative}

\noindent\textbf{Downstream classification.}
Tab.~\ref{tab:cls_results} shows the classification results. Our method achieves the best bAcc and macro-F1 on all three datasets (MIMIC-LT: $0.162$; CT-RATE: $0.193$; NIH-LT: $0.107$) and is the only augmentation strategy that consistently improves over the non-augmented baseline across both modalities. Vanilla FM and CBFM can hurt classification on MIMIC-LT, while CPD improves on MIMIC-LT but is less consistent across datasets. Our method also exhibits the lowest variance across seeds, suggesting more stability. Qualitative examples in Fig.~\ref{fig:qualitative} illustrate that our method produces visually coherent tail-class samples while preserving intra-subclass diversity.

\noindent\textbf{Ablation.}
Tab.~\ref{tab:cls_ablation_mimic_ctrate} isolates the two pipeline stages. On MIMIC-LT, optimization alone lifts bAcc from $0.149$ to $0.158$, while subclassing (with or without optimization) reaches $0.162$. On CT-RATE, optimization without subclassing degrades performance ($0.173$) by averaging over conflicting modes, whereas the full pipeline recovers and improves the baseline ($0.193$). On NIH-LT, gains arise only from combining both stages, improving from $0.098$ to $0.107$.

\noindent\textbf{Generative metrics.}
Tab.~\ref{tab:gen_metrics} shows the generative metrics.
On MIMIC-LT our method yields the best FID ($0.045$) and IRS ($0.668$), indicating simultaneous gains in fidelity and diversity. On CT-RATE, CBFM achieves a marginally better FID ($27.00$ vs.\ $29.06$), yet IRS scores similar ($0.713$ vs.\ $0.712$). With its lower classification performance on MIMIC-LT, this suggests that CBFM's FID gain on CT-RATE likely reflects memorization of rare samples rather than true distributional coverage.
On NIH-LT, we obtain the best FID ($0.062$), which paired with the classifier results, suggests improved fidelity.

\noindent\textbf{Subclass analysis.}
To verify that induced subclasses capture pathology-relevant structure rather than acquisition metadata, we train metadata-only multinomial logistic regression to predict $k$, using grouped stratified 5-fold cross-validation with \texttt{subject\_id} grouping to prevent leakage. Tab.~\ref{tab:confounder_probe_mimic} shows that confounders are detectable but not dominant: the macro $\Delta$bAcc over a permutation baseline is only $0.038$, and 5 of 13 evaluated classes are at permutation-level performance (NearChance $= 0.385$). Tab.~\ref{tab:kknn_subclass_vs_random} further confirms non-random subclass structure: kNN purity of learned subclasses exceeds matched random partitions by $+0.164$ (CXR DenseNet) and $+0.100$ (BioViL), measured in two embedding spaces.

\noindent\textbf{Discussion.}
%Our subclass priors consistently improve classification while maintaining or improving generative diversity, confirming that partitioning heterogeneous medical conditionals and localizing source distributions yields more informative synthetic data for tail classes. The ablation demonstrates that both stages are necessary: subclassing resolves modal ambiguity that directional optimization alone cannot address, while source optimization reduces the residual within-subclass dispersion that subclassing alone leaves intact. 
Subclasses are induced in a learned latent space and may partially reflect acquisition-related variation alongside pathology; our confounder probes show this effect is present but not dominant. Extremely small tail subclasses can yield unstable partitions, which we mitigate via EBIC model selection with a conservative $\gamma$ and minimum-size constraints. The offline nature of our approach (under one GPU-hour per dataset) makes it straightforward to integrate into existing flow-matching pipelines without modifying the training loop.

\section{Conclusion}

We introduced a subclass-conditional flow-matching framework for long-tailed medical image generation that partitions heterogeneous labels, learns subclass-dependent sources, and enforces geometric control of transport. This principled, fully offline strategy reduces conditional ambiguity and stabilizes trajectories at negligible cost. Across chest X-ray and CT benchmarks, it consistently improves tail fidelity, diversity, and downstream balanced accuracy over strong baselines, highlighting source design and conditional refinement as effective levers for robust generative augmentation in clinical long-tail settings.

%We introduced informative subclass priors for conditional flow matching in long-tailed medical image generation: GMM-induced subclass assignments partition heterogeneous conditionals into coherent submodes, subclass-dependent sources shorten and stabilize transport paths, and geometry regularization prevents degenerate solutions. On chest X-ray and CT benchmarks, this fully offline strategy is a promising augmentation approach that reliably improves downstream classification over a non-augmented baseline across both modalities, at negligible cost. %We hope this work encourages further exploration of source distribution design for conditional generation in medical imaging and beyond.

\begin{credits}
\subsubsection{\ackname} We acknowledge HPC resources from NHR@FAU (projects b143dc, b180dc), funded by federal and Bavarian state authorities and Gerhard Wellein's HPC approach. NHR@FAU hardware is partially funded by DFG 440719683. Additional support was received from ERC projects MIA-NORMAL 101083647, DFG 513220538 and 512819079, and the state of Bavaria (HTA and the Bavarian Foundation Model Initiative). We further acknowledge resources provided by the Isambard-AI National AI Research Resource (AIRR), operated by the University of Bristol and funded by DSIT via UKRI and STFC [ST/AIRR/I-A-I/1023]~\cite{mcintoshsmith2024isambardai}. We were supported by coding agents and LLMs from Anthropic, OpenAI, Google, and Mistral AI, for text polishing, coding, experiment orchestration, and cluster monitoring.

\subsubsection{\discintname}
The authors have no relevant competing interests.
\end{credits}

%\newpage
 \bibliographystyle{splncs04}
 \bibliography{main}

\end{document}

%% file: figures/method.tex
\begin{figure}[t]
    \centering
    \includegraphics[width=\linewidth]{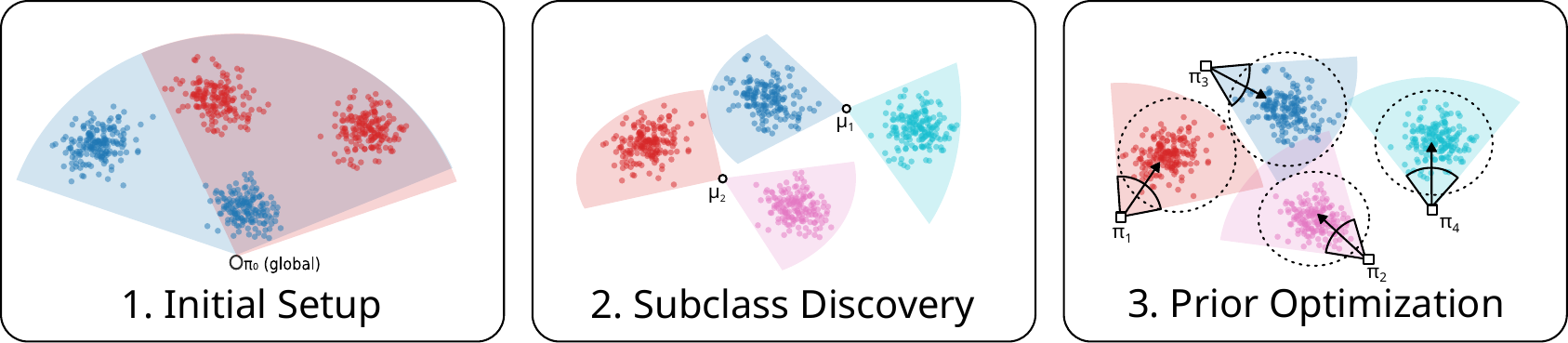}
    \caption{\textbf{Left:} Coarse conditionings and a single source yield a large variance of conditional probability paths. \textbf{Middle:} We obtain finer conditionings by fitting a mixture on the residual directions from the class centers. \textbf{Right:} We obtain better directional alignment by assigning a source to each subclass and optimizing them to a common direction, each bounded by a radial cap.}
    \label{fig:method}
\end{figure}

%% file: tables/diagnostics_unified.tex
% ============================================================
%  Unified diagnostics block
%  Row 1: Table 1 (split summary) | Table 2 (geometric diagnostics)
%  Row 2: Figure (balance)        | Table 3 (confounder probe)
% ============================================================
\begin{table}[t]
\centering
\makeatletter

% ---- ROW 1: Table 1 + Table 2 side by side ----
\begin{minipage}[t]{0.54\columnwidth}
  \vspace{0pt}
  \centering
  \def\@captype{table}
  \caption{Split summary based on the EBIC criterion.
  $B$ = \#base classes, $K$ = \#subclasses.
  ``$\overline{\text{split}}$'' means $K_c{=}1$; ``split'' means $K_c{>}1$.}
  \label{tab:subclass}
  \resizebox{\linewidth}{!}{%
    \begin{tabular}{lcccccc}
      \toprule
      Dataset & $B$ & $K$ & \#$\overline{\text{split}}$ & $\lvert\overline{\text{split}}\textsubscript{max}\rvert$ & $\lvert split\textsubscript{min}\rvert$ & $\min n_{c,k}\!\in\!\text{split}$\\
      \midrule
      MIMIC-LT & 19 & 91 & 6 & 254 & 609 & 106 \\
      NIH-LT & 20 & 80 & 6 & 213 & 229 & 95 \\
      CT-RATE  & 15 & 53 & 7 & 244 & 335 & 97 \\
      \bottomrule
    \end{tabular}%
  }
\end{minipage}%
\hfill
\begin{minipage}[t]{0.42\columnwidth}
  \vspace{0pt}
  \centering
  \def\@captype{table}
  \caption{Geometric diagnostics (before $\rightarrow$ after source optimization). }
  \label{tab:geom_diag}
  \resizebox{\linewidth}{!}{%
    \begin{tabular}{lcc}
      \toprule
      Dataset & 
      \shortstack{$\cos_{\mathrm{mean},w} \uparrow$ \\ [-0.6ex] {\tiny directional alignment}} & 
      \shortstack{$r_{\mathrm{rel},w} \downarrow$ \\ [-0.6ex] {\tiny radial spread}} \\
      \midrule
      MIMIC-LT & $0.00 \rightarrow 0.68$ & $0.08 \rightarrow 0.04$ \\
      NIH-LT   & $0.00 \rightarrow 0.68$ & $0.09 \rightarrow 0.05$ \\
      CT-RATE  & $0.00 \rightarrow 0.69$ & $0.07 \rightarrow 0.03$ \\
      \bottomrule
    \end{tabular}%
  }
\end{minipage}

\vspace{0.8em}

% ---- ROW 2: Figure + Table 3 side by side ----
\begin{minipage}[t]{0.48\columnwidth}
  \vspace{0pt}
  \centering
  \def\@captype{figure}
  \includegraphics[width=\linewidth]{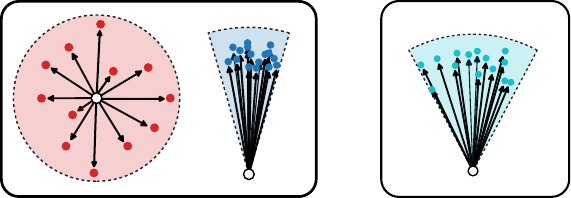}
  \caption{\textbf{Left:} Close sources produce noisy directions; too distant sources produce entangled paths.
  \textbf{Right:} We balance angular concentration and distance to targets.}
  \label{fig:bal_geo}
\end{minipage}%
\hfill
\begin{minipage}[t]{0.50\columnwidth}
  \vspace{0pt}
  \centering
  \def\@captype{table}
  \caption{Aggregate confounder-probe results on MIMIC-LT subclasses.
  Macro averages are unweighted over base classes ($n{=}13$);
  weighted averages use per-class sample counts.
  NearChanceShare: fraction of classes with $\Delta$bAcc ${\le}\,0.03$.}
  \label{tab:confounder_probe_mimic}
  \resizebox{\linewidth}{!}{%
    \begin{tabular}{lcccc}
      \toprule
      Aggregation & bAcc & PermbAcc & $\Delta$bAcc & NearChance \\
      \midrule
      Macro (over labels)   & 0.273 & 0.235 & 0.038 & 0.385 \\
      Weighted (by samples) & 0.157 & 0.117 & 0.040 & 0.385 \\
      \bottomrule
    \end{tabular}%
  }
\end{minipage}

\makeatother
\end{table}

%% file: tables/table4-7.tex
% ==========================================
% Table A: Classifier results (bAcc, p, F1)
% ==========================================
\begin{table}[t]
\centering
\setlength{\tabcolsep}{3.2pt}
\caption{Downstream \emph{classification} on MIMIC-LT, CT-RATE, and NIH-LT. Best per column in \textbf{bold}, second best \underline{underlined}. Two-sided Wilcoxon signed-rank $p$-values on bAcc compare \textbf{Ours} to each baseline (paired over nine runs).}
\label{tab:cls_results}
\resizebox{\textwidth}{!}{%
\begin{tabular}{l ccc ccc ccc}
\toprule
& \multicolumn{3}{c}{MIMIC-LT} & \multicolumn{3}{c}{CT-RATE} & \multicolumn{3}{c}{NIH-LT} \\
\cmidrule(lr){2-4}\cmidrule(lr){5-7}\cmidrule(lr){8-10}
Method
  & bAcc$\uparrow$ & $p$ (bAcc) & F1$\uparrow$
  & bAcc$\uparrow$ & $p$ (bAcc) & F1$\uparrow$
  & bAcc$\uparrow$ & $p$ (bAcc) & F1$\uparrow$ \\
\midrule
Real (no aug.)
  & $0.157\!_{\pm0.014}$ & -- & $0.158\!_{\pm0.013}$
  & $0.176\!_{\pm0.009}$ & -- & $0.164\!_{\pm0.012}$
  & $0.098\!_{\pm0.028}$ & -- & $0.096\!_{\pm0.035}$ \\
Vanilla FM
  & $0.149\!_{\pm0.008}$ & $0.0039$ & $0.153\!_{\pm0.008}$
  & \underline{$0.191\!_{\pm0.012}$} & $1.0000$ & $0.171\!_{\pm0.023}$
  & \underline{$0.098\!_{\pm0.028}$} & $0.5703$ & \underline{$0.096\!_{\pm0.035}$} \\
CBFM~\cite{qin2023cb}
  & $0.150\!_{\pm0.009}$ & $0.0039$ & $0.153\!_{\pm0.009}$
  & $0.179\!_{\pm0.017}$ & $0.0251$ & $0.165\!_{\pm0.019}$
  & $0.087\!_{\pm0.025}$ & $0.1641$ & $0.078\!_{\pm0.031}$ \\
CPD~\cite{issachar2025designing}
  & \underline{$0.159\!_{\pm0.013}$} & $0.5527$ & \underline{$0.159\!_{\pm0.017}$}
  & $0.189\!_{\pm0.012}$ & $0.4961$ & \underline{$0.173\!_{\pm0.014}$}
  & $0.091\!_{\pm0.029}$ & $0.2031$ & $0.088\!_{\pm0.037}$ \\
\textbf{Ours}
  & $\mathbf{0.162\!_{\pm0.005}}$ & -- & $\mathbf{0.163\!_{\pm0.006}}$
  & $\mathbf{0.193\!_{\pm0.014}}$ & -- & $\mathbf{0.177\!_{\pm0.011}}$
  & $\mathbf{0.107\!_{\pm0.022}}$ & -- & $\mathbf{0.107\!_{\pm0.033}}$ \\
\bottomrule
\end{tabular}
}%
\end{table}

% ==========================================
% Table B: Generative metrics (FID, IRS)
% ==========================================
\begin{table}[t]
\centering
\setlength{\tabcolsep}{3.2pt}
\caption{Generative metrics on MIMIC-LT, CT-RATE, and NIH-LT (mean$\pm$std over three seeds). Best per metric column in \textbf{bold}, second best \underline{underlined}.}
\label{tab:gen_metrics}
\resizebox{\textwidth}{!}{%
\begin{tabular}{l cc cc cc}
\toprule
& \multicolumn{2}{c}{MIMIC-LT} & \multicolumn{2}{c}{CT-RATE} & \multicolumn{2}{c}{NIH-LT} \\
\cmidrule(lr){2-3}\cmidrule(lr){4-5}\cmidrule(lr){6-7}
Method
  & FID$\downarrow$ & IRS$\uparrow$
  & FID$\downarrow$ & IRS$\uparrow$
  & FID$\downarrow$ & IRS$\uparrow$ \\
\midrule
Vanilla FM
  & $0.048\!_{\pm0.002}$ & $0.615\!_{\pm0.017}$
  & \underline{$27.11\!_{\pm1.25}$} & $0.694\!_{\pm0.022}$
  & $0.067\!_{\pm0.001}$ & $\mathbf{0.680\!_{\pm0.020}}$ \\
CBFM~\cite{qin2023cb}
  & \underline{$0.047\!_{\pm0.003}$} & \underline{$0.626\!_{\pm0.020}$}
  & $\mathbf{27.00\!_{\pm0.94}}$ & $\mathbf{0.713\!_{\pm0.019}}$
  & \underline{$0.065\!_{\pm0.001}$} & \underline{$0.649\!_{\pm0.008}$} \\
CPD~\cite{issachar2025designing}
  & $0.052\!_{\pm0.003}$ & $0.564\!_{\pm0.007}$
  & $28.48\!_{\pm0.70}$ & $0.671\!_{\pm0.039}$
  & $0.072\!_{\pm0.002}$ & $0.631\!_{\pm0.004}$ \\
\textbf{Ours}
  & $\mathbf{0.045\!_{\pm0.000}}$ & $\mathbf{0.668\!_{\pm0.011}}$
  & $29.06\!_{\pm0.36}$ & \underline{$0.712\!_{\pm0.022}$}
  & $\mathbf{0.062\!_{\pm0.001}}$ & $0.555\!_{\pm0.052}$ \\
\bottomrule
\end{tabular}
}%
\end{table}

\begin{table}[t]
\centering
\scriptsize
\setlength{\tabcolsep}{4pt}
\makeatletter

% ---- ROW 1: Ablation (left) + kNN purity (right) ----
\begin{minipage}[t]{0.49\linewidth}
  \vspace{0pt}\centering
  \def\@captype{table}
  \caption{Ablation: bAcc (mean$_{\pm\text{std}}$ over 9 seeds). Both stages contribute; their combination is best.}
  \label{tab:cls_ablation_mimic_ctrate}
  \resizebox{\linewidth}{!}{%
  \begin{tabular}{cc ccc}
  \toprule
  Sub. & Opt. & MIMIC-LT & CT-RATE & NIH-LT \\
  \midrule
   &  & $0.149\!_{\pm0.008}$ & $\underline{0.191\!_{\pm0.012}}$ & $0.098\!_{\pm0.028}$ \\
   & \checkmark & $0.158\!_{\pm0.007}$ & $0.173\!_{\pm0.016}$ & $0.098\!_{\pm0.030}$ \\
  \checkmark &  & $\mathbf{0.162\!_{\pm0.006}}$ & $0.182\!_{\pm0.017}$ & $\underline{0.099\!_{\pm0.023}}$ \\
  \checkmark & \checkmark & $\mathbf{0.162\!_{\pm0.005}}$ & $\mathbf{0.193\!_{\pm0.014}}$ & $\mathbf{0.107\!_{\pm0.022}}$ \\
  \bottomrule
  \end{tabular}
  }%
\end{minipage}%
\hfill
\begin{minipage}[t]{0.49\linewidth}
  \vspace{0pt}\centering
  \def\@captype{table}
  \caption{Weighted kNN purity: learned subclasses vs.\ matched random partitions in two independent embeddings.}
  \label{tab:kknn_subclass_vs_random}
  \resizebox{\linewidth}{!}{%
  \begin{tabular}{l ccc}
  \toprule
  Embedding & Learned$\uparrow$ & Random & $\Delta\!\uparrow$ \\
  \midrule
  CXR DenseNet~\cite{dombrowski2026pso} & 0.295 & 0.131 & 0.164 \\
  BioViL~\cite{boecking2022biovil} & 0.231 & 0.131 & 0.100 \\
  \bottomrule
  \end{tabular}
  }%
\end{minipage}

\makeatother
\end{table}

%% file: figures/qualitative.tex
\begin{figure*}[t]
\centering
\resizebox{\textwidth}{!}{%
\begin{tikzpicture}[
  imgnode/.style={draw=black!40, thick, minimum width=2.1cm, minimum height=2.1cm,
                  inner sep=0pt, rounded corners=2pt, font=\scriptsize\sffamily,
                  align=center, fill=gray!8},
  lbl/.style={font=\scriptsize\sffamily, text=black!70},
  grouplbl/.style={font=\small\sffamily\bfseries},
]

% helper for image placeholders from images/ folder
\newcommand{\imgph}[1]{\includegraphics[width=2.1cm,height=2.1cm]{images/#1}}

% --- Main Grid Column labels ---
\node[grouplbl] at (0, 1.4) {Real};
\node[grouplbl] at (2.2, 1.4) {Vanilla FM};
\node[grouplbl] at (4.4, 1.4) {CBFM};
\node[grouplbl] at (6.6, 1.4) {Ours};

% --- Diversity Column label ---
\node[grouplbl] at (10.1, 1.4) {Ours (Diversity)};

% --- Subtle vertical divider ---
\draw[draw=black!20, thick, dashed] (7.9, 1.7) -- (7.9, -5.6);

% --- Row 1: MIMIC-LT tail class ---
\node[lbl, rotate=90, anchor=south] at (-1.3, 0) {Pneumothorax};
\node[imgnode] (r1c1) at (0, 0) {\imgph{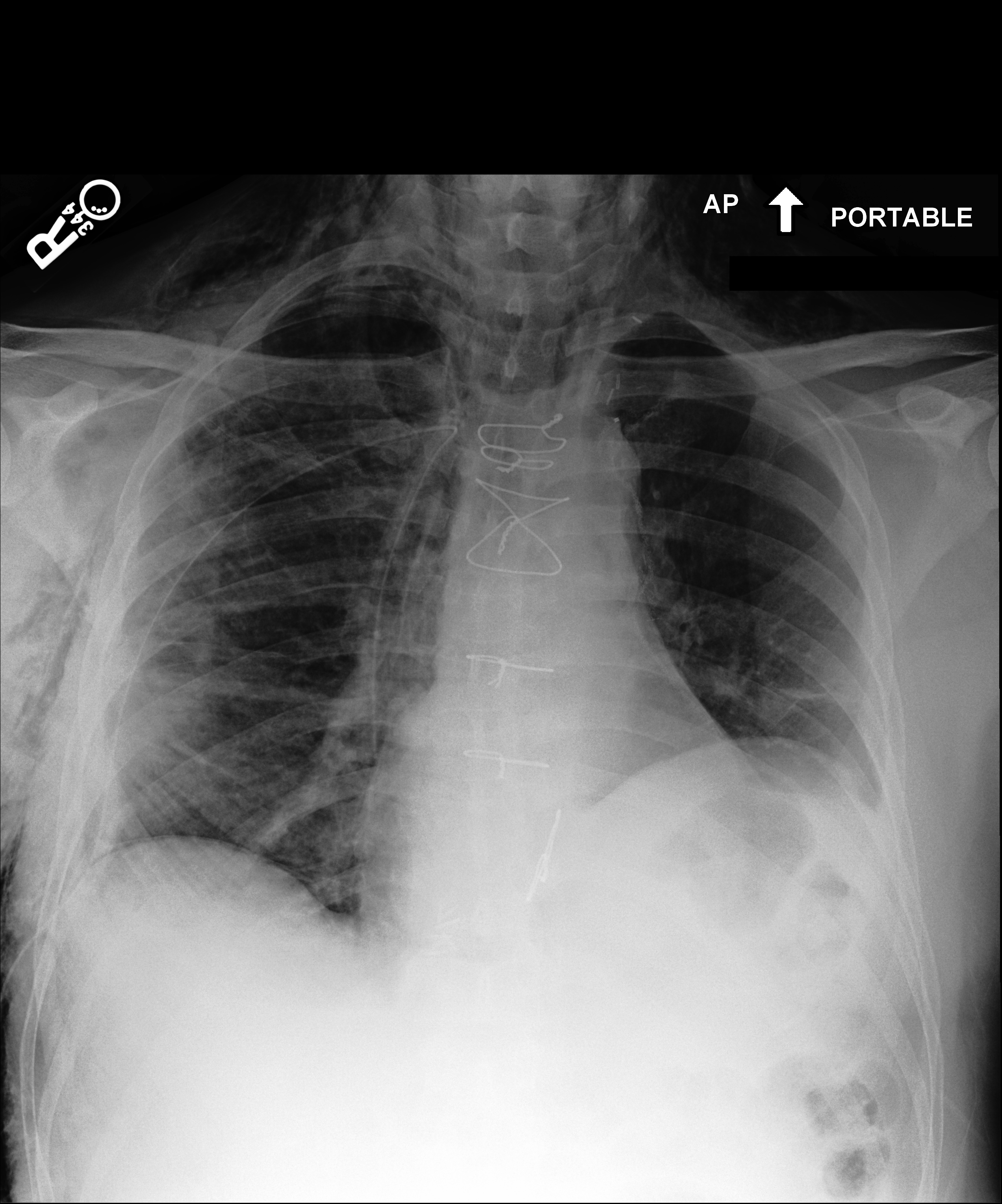}};
\node[imgnode] (r1c2) at (2.2, 0) {\imgph{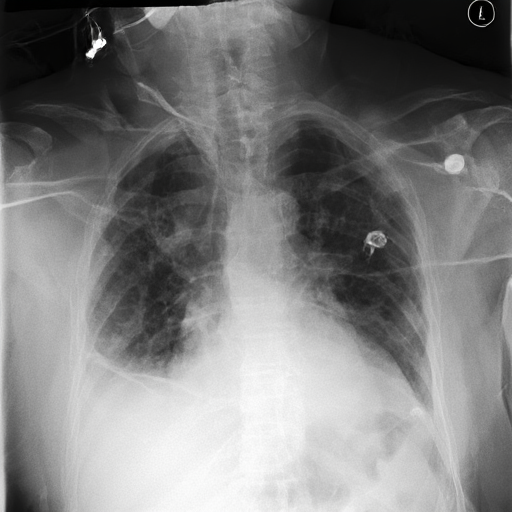}};
\node[imgnode] (r1c3) at (4.4, 0) {\imgph{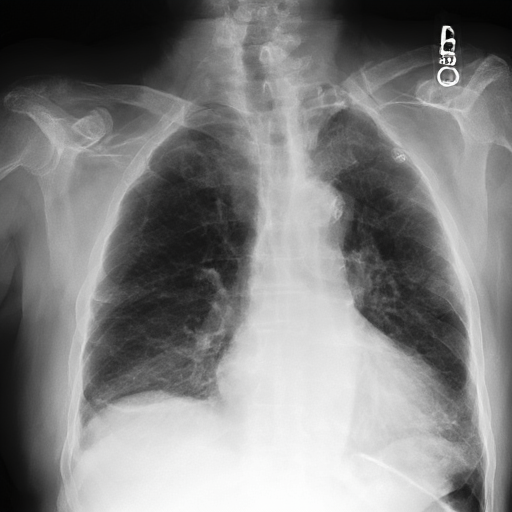}};
\node[imgnode] (r1c4) at (6.6, 0) {\imgph{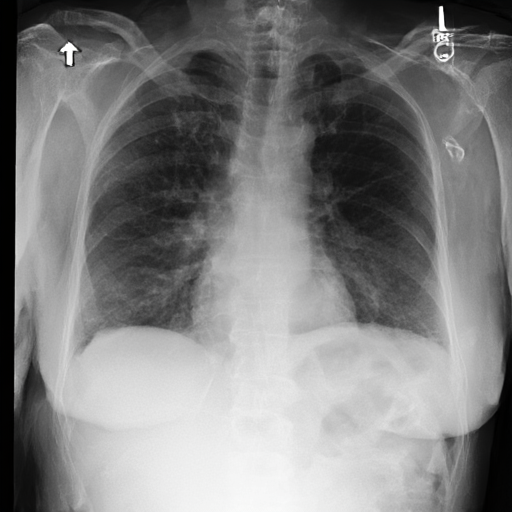}};

% --- Row 2: NIH-LT tail class ---
\node[lbl, rotate=90, anchor=south] at (-1.3, -2.2) {Pneumomediastinum};
\node[imgnode] (r2c1) at (0, -2.2) {\imgph{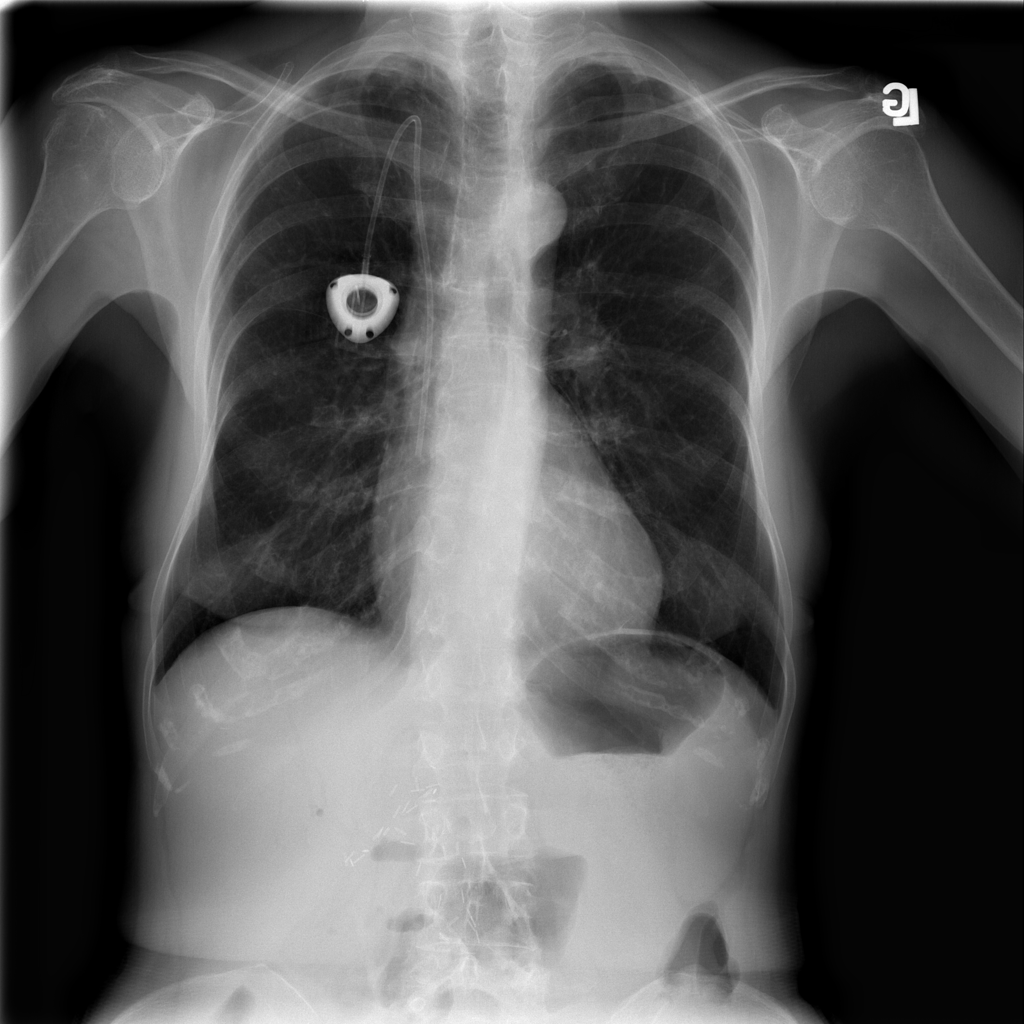}};
\node[imgnode] (r2c2) at (2.2, -2.2) {\imgph{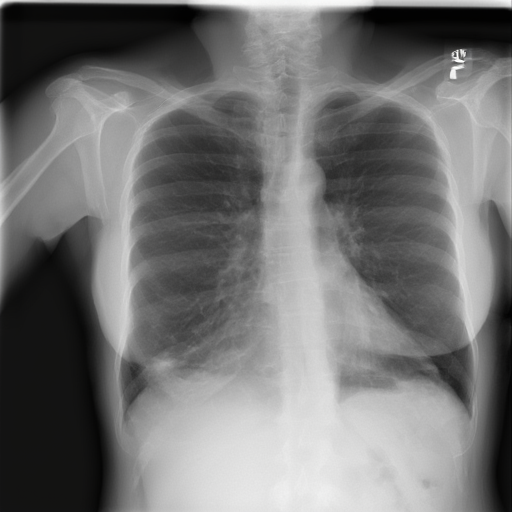}};
\node[imgnode] (r2c3) at (4.4, -2.2) {\imgph{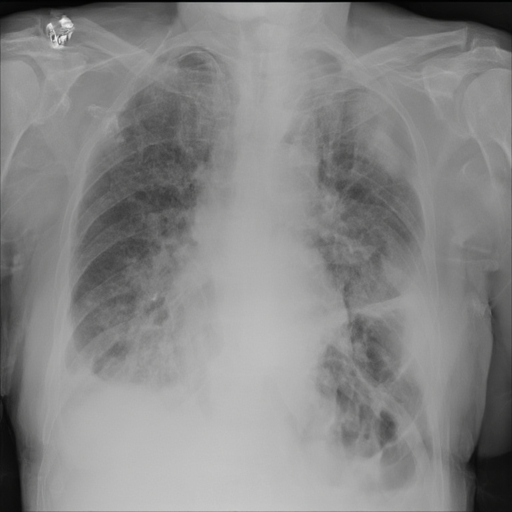}};
\node[imgnode] (r2c4) at (6.6, -2.2) {\imgph{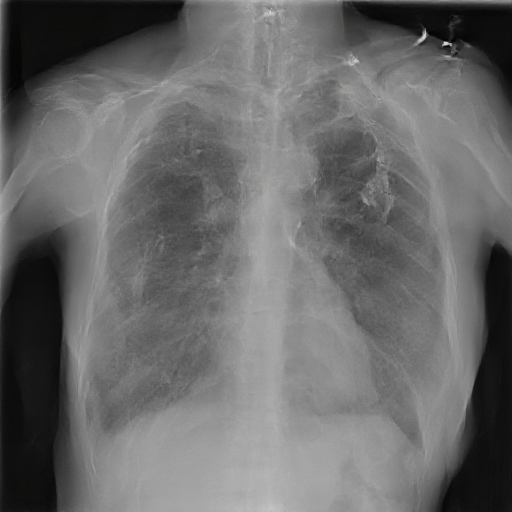}};

% --- Row 3: CT-RATE tail class ---
\node[lbl, rotate=90, anchor=south] at (-1.3, -4.4) {Bronchiectasis};
\node[imgnode] (r3c1) at (0, -4.4) {\imgph{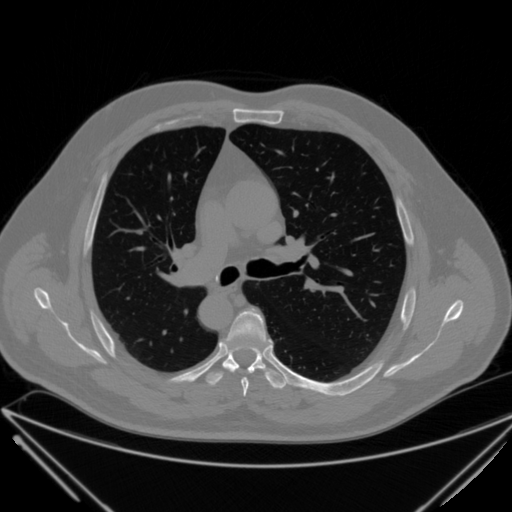}};
\node[imgnode] (r3c2) at (2.2, -4.4) {\imgph{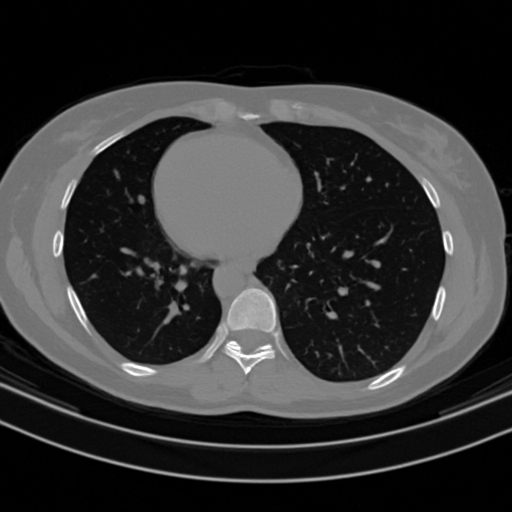}};
\node[imgnode] (r3c3) at (4.4, -4.4) {\imgph{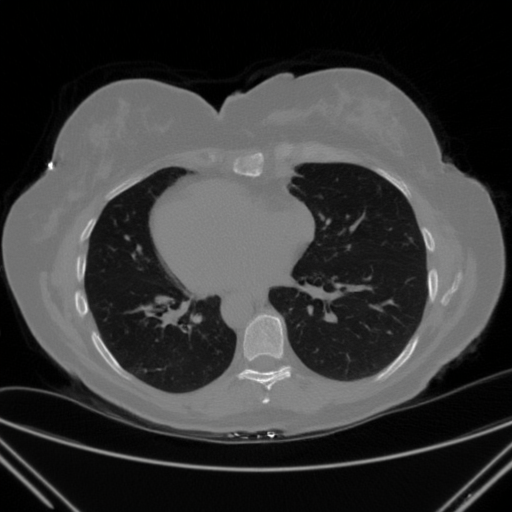}};
\node[imgnode] (r3c4) at (6.6, -4.4) {\imgph{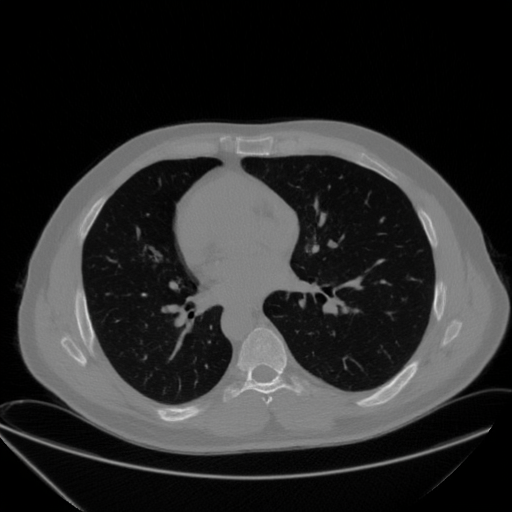}};
% --- Diversity block (Right side, 2x3 arrangement) ---
\node[imgnode] (d1) at (9.0, 0) {\imgph{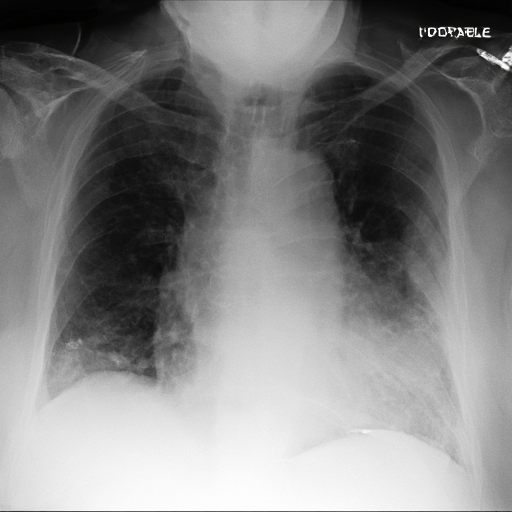}};
\node[imgnode] (d2) at (11.2, 0) {\imgph{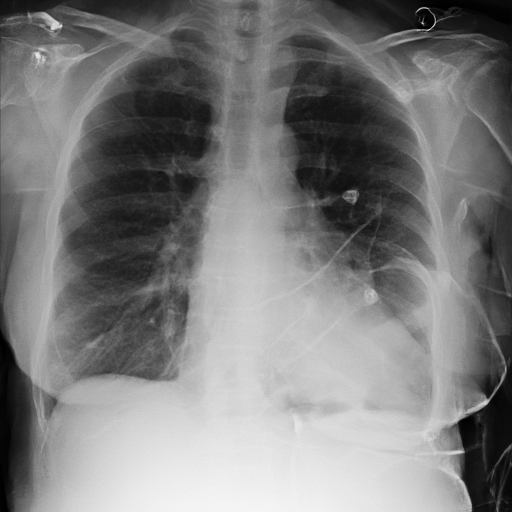}};
\node[imgnode] (d3) at (9.0, -2.2) {\imgph{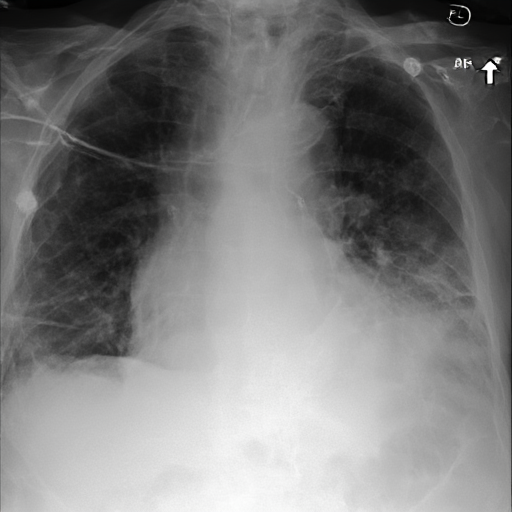}};
\node[imgnode] (d4) at (11.2, -2.2) {\imgph{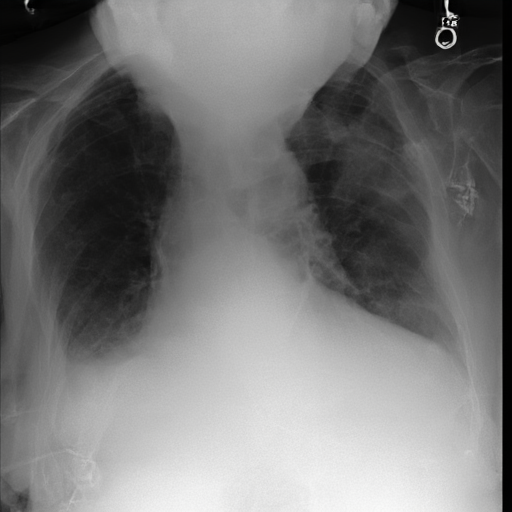}};
\node[imgnode] (d5) at (9.0, -4.4) {\imgph{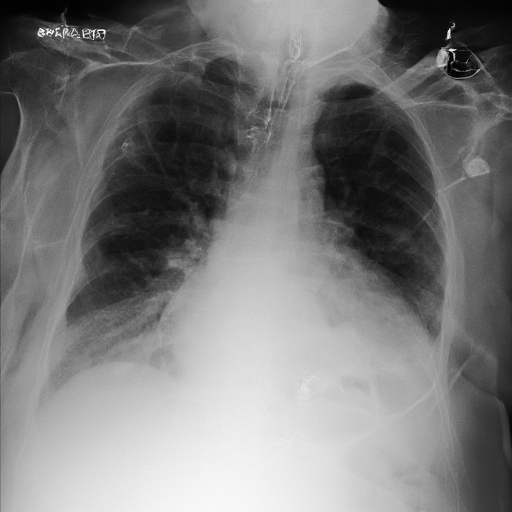}};

% Fill the 6th slot with the description text to perfectly utilize the space
\node[align=center, font=\scriptsize\sffamily, text width=2.1cm, text=black!80]
  at (11.2, -4.4) {\textbf{Intra-subclass}\\ \textbf{diversity}\\ (Pneumothorax)};

\end{tikzpicture}
}%
\caption{Qualitative comparison of real and generated tail-class images. Row 1: MIMIC-LTtail class; Row 2: NIH-LT tail class; Row 3: CT-RATE tail class. Right: five random draws from a single subclass of our method, illustrating intra-subclass diversity.}
\label{fig:qualitative}
\end{figure*}